\documentclass[pra,aps,twocolumn,showpacs,preprintnumbers,amsmath,amssymb]{revtex4}
\usepackage{dcolumn}
\usepackage{bm}
\usepackage{amsfonts,latexsym,graphicx}

\newcommand{\ket}[1]{\displaystyle{|#1\rangle}}

\newcommand{\om}{\omega}

\newcommand{\si}{\sigma}

\begin{document}
\title{Connection among entanglement, mixedness and nonlocality in a dynamical context}

\author{Laura Mazzola$^1$}
\email{laura.mazzola@utu.fi}
\author{Bruno Bellomo$^2$}
\author{Rosario Lo Franco$^2$}
\email{lofranco@fisica.unipa.it}
\author{Giuseppe Compagno$^2$}
\affiliation{$^1$Turku Centre for Quantum Physics, Department of
Physics and Astronomy, University of Turku, FI-20014 Turun
yliopisto, Finland\\$^2$CNISM and Dipartimento di Scienze Fisiche ed
Astronomiche, Universit\`{a} di Palermo, via Archirafi 36, 90123
Palermo, Italy}

\date{\today}

\begin{abstract}
We investigate the dynamical relations among entanglement, mixedness
and nonlocality, quantified by concurrence $C$, purity $P$ and
maximum of Bell function $B$, respectively, in a system of two
qubits in a common structured reservoir. To this aim we introduce
the $C$-$P$-$B$ parameter space and analyze the time evolution of
the point representative of the system state in such a space. The
dynamical interplay among entanglement, mixedness and nonlocality
strongly depends on the initial state of the system. For a
two-excitation Bell state the representative point draws a
multi-branch curve in the $C$-$P$-$B$ space and we show that a
closed relation among these quantifiers does not hold. By extending
the known relation between $C$ and $B$ for pure states, we give an
expression among the three quantifiers for mixed states. In this
equation we introduce a quantity, vanishing for pure states which
has not in general a closed form in terms of $C$, $P$ and $B$.
Finally we demonstrate that for an initial one-excitation Bell state
a closed $C$-$P$-$B$ relation instead exists and the system evolves
remaining always a maximally entangled mixed state.
\end{abstract}

\pacs{03.67.Bg, 03.65.Yz, 42.40.-p, 03.65.Ud}

\maketitle

\section{Introduction \label{sec-intro}}
Entanglement, mixedness and nonlocality are among the main
properties describing the quantum features of a composite system.
Entanglement is linked to quantum correlations
\cite{horodecki2009RMP} and for a two-qubit state can be quantified,
e.g., by concurrence $C$ \cite{wootters1998PRL}, while for a
multipartite system its characterization remains an open problem.
Mixedness, namely how much the state of a quantum system is far from
being pure, can be quantified by the purity $P$ (linked to the
linear entropy) or by the Von Neumann entropy \cite{nielsenchuang}.
Nonlocality describes the part of quantum correlations which cannot
be reproduced by any classical local model \cite{bell}. It is
typically characterized by combination of correlations averages,
named Bell function, violating some Bell inequality \cite{clauser}.
The value obtained for the Bell function depends on the state of the
system and on some parameters determined by the experimental
settings. It may happen that, for some of these settings, the value
obtained for the Bell function does not violate the Bell inequality.
It is therefore appropriate, in general, to fix the external
parameters to obtain the maximum possible value $B$ for the Bell
function. In this sense, the maximum of the Bell function $B$
individuates at best the presence of nonlocality
\cite{horodecki1995PLA}. All of these quantifiers may be obtained by
measurements on the system. The properties they represent play an
important role in quantum information science, such as in the
realization of device-independent and security-proof quantum key
distribution protocols
\cite{nielsenchuang,acin2006PRL,gisin2007natphoton}. In applicative
contexts, it has been shown that also states nonviolating any Bell inequality can be used for teleportation
\cite{Popescu1994PRL} and
that every entangled state shows some hidden nonlocality
\cite{masanes2008PRL} which may be exploited using local filtering
\cite{forster2009PRL}.

The values of the three quantities $C$, $P$ and $B$, are related and
connections among pairs of them have been widely investigated. These
connections are far from being trivial. For example, although for pure
states the presence of entanglement implies nonlocality
\cite{gisin1991PLA}, on the contrary for mixed states a given amount
of entanglement does not necessarily guarantee violation of a Bell
inequality \cite{werner1989PRA,barrett2002PRA,acingisin2006PRA}. In
particular, for bipartite systems, a range of possible Bell
inequality violations corresponds to a certain amount of
entanglement \cite{verstraete2002PRL}, while states with a different
degree of entanglement can violate a Bell inequality of the same
amount \cite{Miranowicz2004}.

The connection between entanglement and mixedness has been
investigated often in the concurrence-purity plane and maximally
entangled mixed two-qubit states for assigned mixedness have been
identified \cite{munro2001PRA}. Their dependence on the quantifiers
has also been pointed out \cite{wei2003PRA}. Moreover, the
entanglement-mixedness relation has been analyzed for some
dynamical systems in the presence of environmental noise
\cite{ziman2005PRA,cardoso2005PRA}.

For what concerns the connection among entanglement, mixedness and
nonlocality, it has been conjectured that the more mixed a system
is, the more entanglement is needed to violate a Bell inequality to
the same amount \cite{munro2001JMO}. However, there are states
having the same amount of entanglement and mixedness but different
values of the Bell function \cite{ghosh2001PRA}. Relations between
entanglement, mixedness and Bell function have been given
analytically for a restrict \cite{derkacz2004PLA},
and numerically for a more general \cite{derkacz2005PRA} class of states. In particular, there are regions of the
concurrence-linear entropy plane where, given concurrence and linear entropy, two families of states
can be discriminated: all states from one family violate the
Clauser-Horne-Shimony-Holt (CHSH) form of Bell inequality while all
states from the other family satisfy it. One may therefore ask if
more general relations involving all these quantities may be put
forward.

Finally, the variety of relations among entanglement, mixedness and
nonlocality in the state space has not yet been examined in a
dynamical context, e.g., by following them in time for a quantum
system interacting with its surroundings. In this case their time
evolution, as characterized by the quantities $C$, $P$ and $B$, can
be rather complex, depending on the structure of the environment and
on the form of the interactions. In fact, typically decay of both
entanglement and nonlocal correlations are expected, even though
revivals or trapping of them may occur as a consequence of memory
effects \cite{bellomo2007PRL,bellomo2008trapping} and/or of interactions
among parts of the system \cite{tanas2006PRA}. On the contrary,
mixedness typically increases during the evolution tending to
different asymptotic values.

The aim of this paper is to investigate the possible connections
among quantifiers $C$, $P$, and $B$ in a dynamical context, and discuss them for a
wide class of two-qubit states. To this purpose we introduce the
three-dimensional $C$-$P$-$B$ parameter space as a tool to analyze
the dynamics of these relations, choosing the paradigmatic open
quantum system of two qubits in a common structured reservoir. The
$C$-$P$-$B$ space appears to be particularly suitable to describe
the dynamical richness of entanglement, mixedness and nonlocality
relations in such a system.

\section{Dynamics in $C$-$P$-$B$ space for common reservoir\label{sec-dynamics}}
Here we investigate the complex relation among entanglement,
mixedness and nonlocality in a specific dynamical
context. As said before, we introduce a tool: the
concurrence-purity-Bell function ($C$-$P$-$B$) parameter space. The
state of the system and its evolution are represented, respectively, by a point of this space and the trajectory it draws
with time. To begin with, we give the expressions
of concurrence, purity and Bell function for a wide class of quantum
states.

\subsection{$C$, $P$ and $B$ for X states \label{sec-Xstate}}
Here, we report the dependence of $C$, $P$ and $B$ on the density
matrix elements for the class of two-qubit states whose density
matrix $\hat{\rho}_X$, in the standard computational basis
$\mathcal{B}=\{\ket{1}\equiv\ket{11},\ket{2}\equiv\ket{10},\ket{3}\equiv\ket{01},\ket{4}\equiv\ket{00}\}$,
has a X structure of the kind
\begin{equation}\label{Xstatesdensitymatrix}
   \hat{\rho}_X = \left(
\begin{array}{cccc}
  \rho_{11} & 0 & 0 & \rho_{14}  \\
  0 & \rho_{22} & \rho_{23} & 0 \\
  0 & \rho_{23}^* & \rho_{33} & 0 \\
  \rho_{14}^* & 0 & 0 & \rho_{44} \\
\end{array}
\right).
\end{equation}
This class of states is sufficiently general to include the
two-qubit states most considered both theoretically and
experimentally, like Bell states (pure two-qubit maximally entangled
states) and Werner states (mixture of Bell states with white noise)
\cite{nielsenchuang,bellomo2008PRA,horodecki2009RMP}. Such
a X structure for the density matrix moreover arises in a wide variety of
physical situations
\cite{hagley1997PRL,bose2001,kwiat2001Nature,pratt2004,wang2006}. A
further remarkable aspect of these X states is that, under various
kinds of dynamics, the initial X structure is maintained during the
evolution \cite{bellomo2007PRL,bellomo2008PRA}. In particular, this
is the case for the model we shall investigate hereafter; this
justifies our choice of this class of quantum states.

For X states of Eq.~(\ref{Xstatesdensitymatrix}) concurrence $C$,
equal to 1 for maximally entangled states and to 0 for separable
states, is given by
\begin{eqnarray}\label{concurrenceandK}
&C=2 \mathrm{max}\{0,K_{1},K_{2}\},& \nonumber \\
& K_{1}=|\rho_{14}|-\sqrt{\rho_{22}\rho_{33}},\
K_{2}=|\rho_{23}|-\sqrt{\rho_{11}\rho_{44} }. &
\end{eqnarray}
The purity $P$, equal to 1 for pure states and to $1/4$ for
completely mixed states, results to be
\begin{equation}
P=\mathrm{Tr}\{\rho^2\}=\sum_{i}{\rho_{ii}^2}+2(|\rho_{23}|^2+|\rho_{14}|^2).
\end{equation}
Using the Horodecki criterion
\cite{horodecki1995PLA}, the maximum of Bell
function can be expressed in terms of three functions $u_1$, $u_2$
and $u_3$ of the density matrix elements as
$B=2\sqrt{\mathrm{max}_{j>k}\{u_j+u_k\}}$, where $j,k=1,2,3$. When
$B$ is larger than the classical threshold 2, no classical local
model may reproduce all correlations of these states. The three
functions $u_j$ are \cite{derkacz2005PRA}
\begin{eqnarray}\label{uXstate}
&u_1=4(|\rho_{14}|+|\rho_{23}|)^2,\quad
u_2=(\rho_{11}+\rho_{44}-\rho_{22}-\rho_{33})^2,&\nonumber\\
&u_3=4(|\rho_{14}|-|\rho_{23}|)^2.&
\end{eqnarray}
Being $u_1$ always larger than $u_3$, the maximum of Bell function for X states results to be
\begin{eqnarray}
&B=\mathrm{max}\{B_1, B_2\},&\nonumber\\
&B_1=2\sqrt{u_1+u_2},\ B_2=2\sqrt{u_1+u_3}.&
\end{eqnarray}

\subsection{The model}\label{sec-model}
The paradigmatic system we examine consists of two identical qubits
interacting with a common zero-temperature leaky cavity. The
Hamiltonian of the total system is $H=H_{0}+H_{int}$ with
$(\hslash=1)$
\begin{equation}
   H_{0}=\om_{0}(\si_{+}^{A}\si_{-}^{A}+\si_{+}^{B}\si_{-}^{B})
   +\sum_{k}\om_{k}a_{k}^{\dag}a_{k},\label{H0bare}
\end{equation}
\begin{equation}
   H_{int}=(\si_{+}^{A}+\si_{+}^{B})\sum_{k}g_{k}a_{k}+\mathrm{h.c.}.\label{Hintbare}
\end{equation}
Here, $\sigma^A_{\pm}$ and $\sigma^B_{\pm}$ are, respectively, the
Pauli raising and lowering operators for atoms $A$ and $B$,
$\omega_0$ is the Bohr frequency of the two atoms, $a_k$ and
$a_k^{\dag}$ are the annihilation and creation operators for the
field mode $k$, and mode $k$ is characterized by the frequency
$\omega_k$ and the coupling constant $g_k$. Since the atoms are
identical and equally coupled to the reservoir, the dynamics of the
two qubits can be effectively described by a four-state system in
which the three states of the triplet, $\ket{00}$, the super-radiant
state $\ket{+}=(\ket{10}+\ket{01})/\sqrt{2}$ and $\ket{11}$, are
coupled to the vacuum in a ladder configuration, and the singlet
state, $\ket{-}=(\ket{10}-\ket{01})/\sqrt{2}$, is completely
decoupled from the other states and from the field
\cite{Mazzola2009ARX}. In particular, the super-radiant state is
coupled to both states $\ket{00}$ and $\ket{11}$ via the
electromagnetic field.

The reservoir is modeled as an infinite sum of harmonic oscillators
and its properties are described through a Lorentzian spectral
distribution
\begin{equation}
   J(\omega)=\frac{1}{2\pi}\frac{\Gamma\lambda^2}{(\om-\om_{0})^2+\lambda^2},
\end{equation}
where the parameter $\lambda$ defines the spectral width of the coupling and $\Gamma$ is related to the decay of the excited state of the qubit in the Markovian limit of flat spectrum (spontaneous emission rate). The ideal cavity limit (no losses) is obtained for $\lambda\rightarrow 0$. The dynamics of this system has been solved exactly (with no perturbation theory or Markov approximation) in Ref.~\cite{Mazzola2009PRA}. Entanglement dynamics has been studied for a large class of initial states in Ref.~\cite{Mazzola2009PRA,Mazzola2009ARX}. Such a system exhibits a rich dynamics due to the memory effects of the non-Markovian environment and the reservoir-mediated interaction between the qubits.

It is thus interesting to investigate the $C$-$P$-$B$ dynamical relation in this physical configuration. The dynamics of the representative point in the $C$-$P$-$B$ parameter space shall allow one to visualize the relations between these three physical quantities. We shall consider initial states with an X form which results to be maintained during the evolutions so that we can use equations of Sec.~\ref{sec-Xstate} to compute $C$, $P$ and $B$. For a given system and fixed initial state the point in the $C$-$P$-$B$ space, representing the state of the system, draws a certain path individuating the dynamical evolution. The flow of time shall be represented by arrows. We shall consider a very narrow Lorentzian distribution to emphasize the memory effects.

\subsection{$\ket{\Psi}$ state dynamics in $C$-$P$-$B$ space}
\begin{figure}
\begin{center}
\includegraphics[width=6.5 cm, height= 4.5 cm]{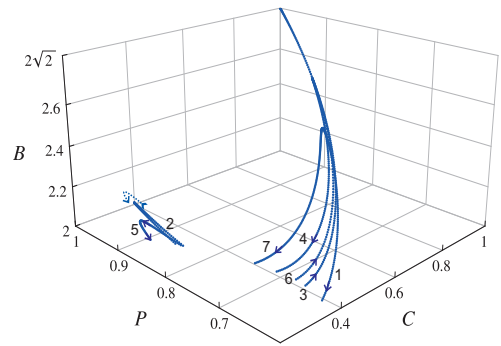}
\end{center}
\caption{\label{fig:3D}\footnotesize (Color online) $C$-$P$-$B$ space curve drawn by the system starting from the initial two-excitation Bell state $\ket{\Psi}$ for $\lambda=10^{-3}\Gamma$. The arrows indicate the time evolution and the numbering from 1-to-7 indicates the different branches (multi-branch behavior) raising from the dynamics. A one-to-one correspondence among the three quantities is not possible here.}
\end{figure}

We start our investigation considering as initial state the two-excitation Bell-state $\ket{\Psi}$
\begin{equation}\label{initial psi state}
   \ket{\Psi}=(\ket{00}+\ket{11})/\sqrt{2},
\end{equation}
whose dynamics is displayed in Fig.~\ref{fig:3D}, where a nontrivial dynamical interplay among $C$, $P$ and $B$ is shown.
Such a plot in the $C$-$P$-$B$ space consists of many branches along which the system moves during the evolution. The separation between the branches depends on the losses of the system. In fact, it can be shown that for a
wider Lorentzian spectral distribution (worse cavity) the branches
become more separated and they reach lower values in the $B$ axis.
Differently, they tend to coincide for a perfect cavity (single mode
reservoir). The trajectory drawn by the system is obtained by
sampling $C$-$P$-$B$ triplets up to a certain time ($200\Gamma t$)
allowing us to bring to light the main features of the dynamics.
Arrows and numbers facilitate the reading of the plot. The state of
the system is initially pure ($P=1$), maximally entangled ($C=1$)
and maximally nonlocal ($B=2\sqrt{2}$). $C$, $P$ and
$B$ deteriorate with time until the representative point has a value of $B$ which satisfies the Bell
inequality (branch 1 of Fig.~\ref{fig:3D}). Now, a
completely new dynamical feature appears: the curve surfaces from
the $B=2$ plane in a region of small concurrence and high purity
(branch 2). This behavior follows from the fact that when the system
is almost pure even a small amount of entanglement induces the
appearance of nonlocality. After such a revival of purity
and nonlocality, the curve sinks again and reappears on the space
region with smaller purity (branch 3). However, the system does not pass through the same
$C$-$P$-$B$ points of the first branch, but it traces a new branch
close to the first one (branch 3). Successively, once again
decoherence effects due to the environment lead to deterioration of $C$,
$P$ and $B$, and a new branch appears (branch 4). The high
non-Markovianity of the reservoir again causes Bell violation on the
high purity/small concurrence region of space (branch 5). The
behavior continues in a similar way and the point draws new branches until a
time after which no violation occurs anymore.

\begin{figure}
{\includegraphics[width=7 cm, height=4.6 cm]{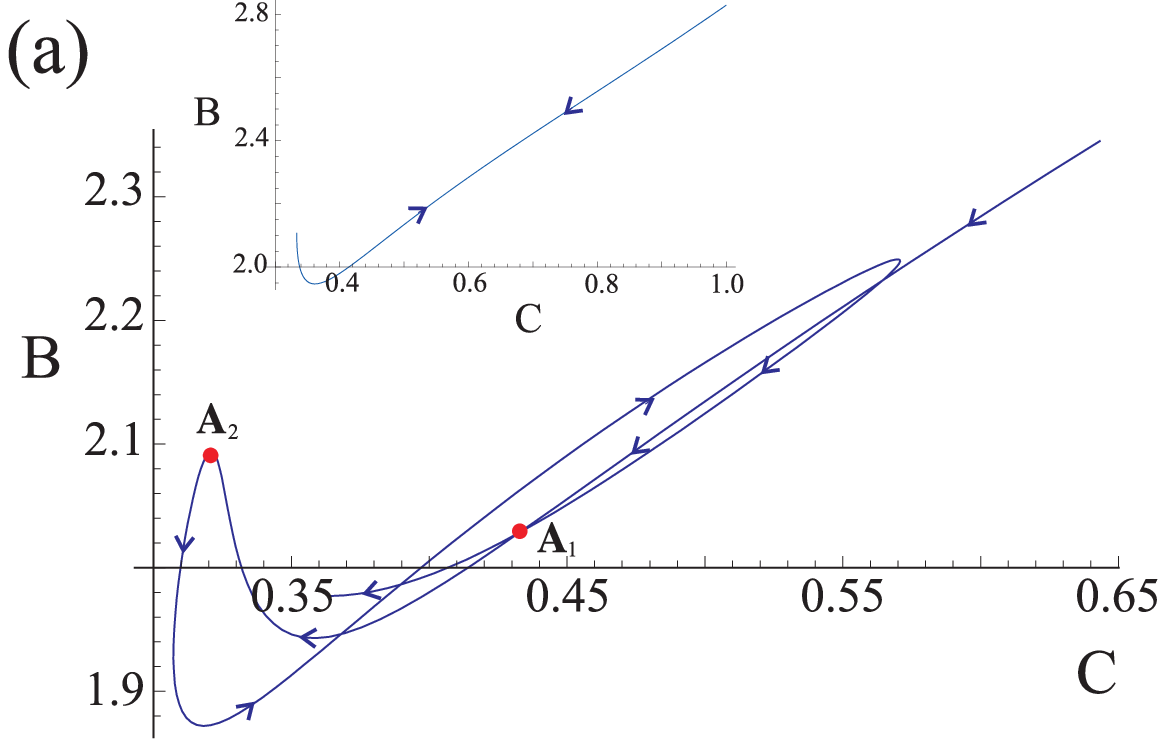}\vspace{0.5 cm}
\includegraphics[width=7 cm, height=4.6 cm]{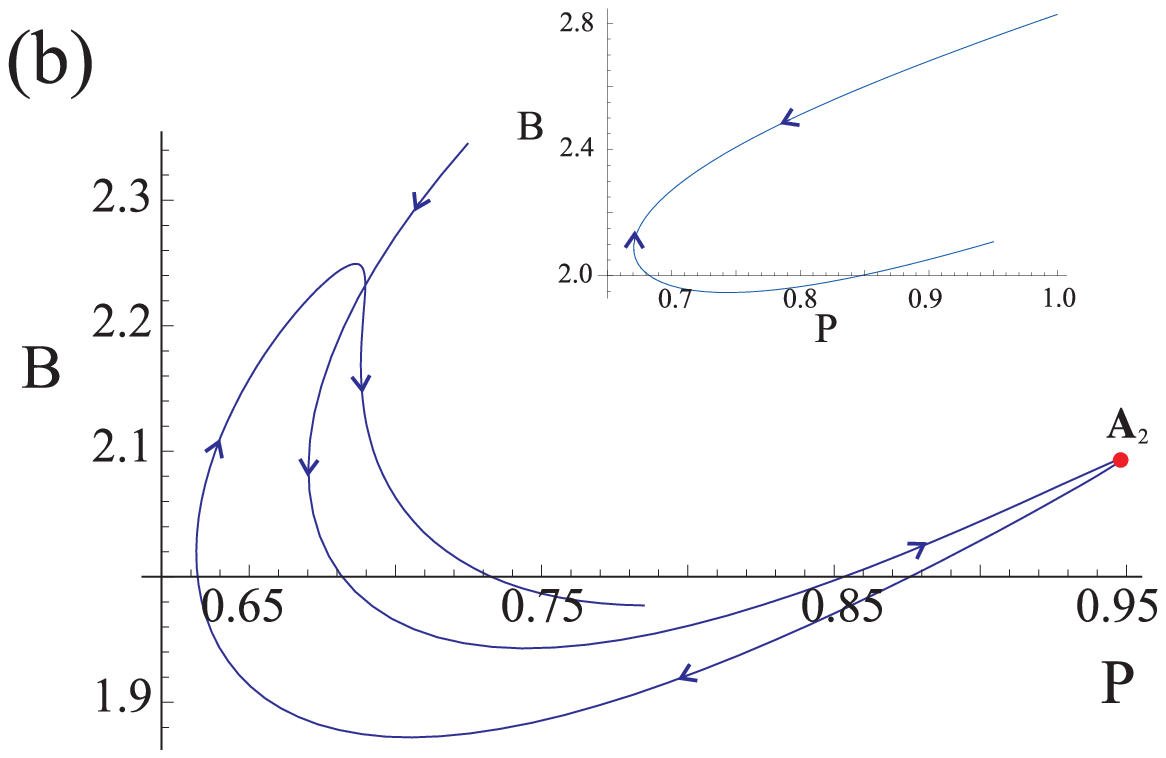}\vspace{0.5 cm}
\includegraphics[width=7 cm, height=5.1 cm]{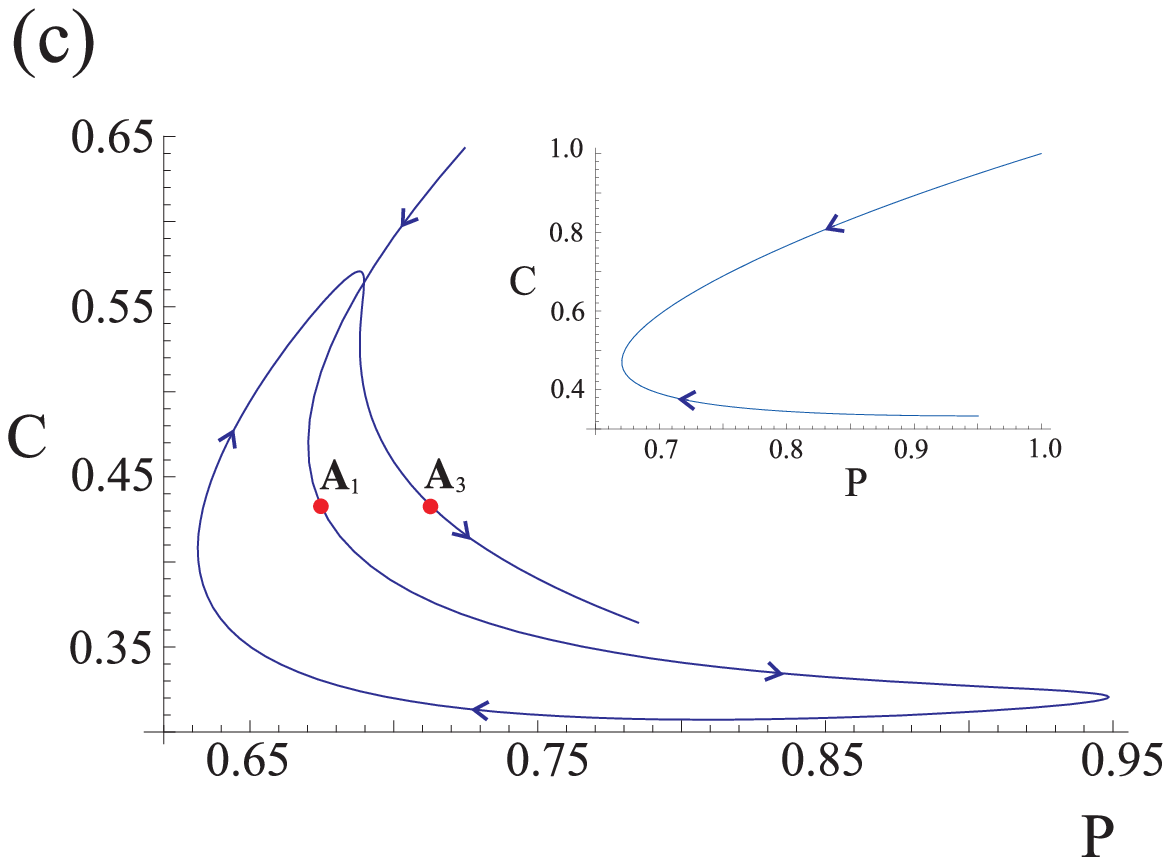}}
\caption{\label{fig:projections}\footnotesize (Color online) Projections of the $C$-$P$-$B$ space curve starting from the two-excitation Bell state $\ket{\Psi}$ on the planes $B$-$C$ (a), $B$-$P$ (b) and $C$-$P$ (c) for $\lambda=10^{-2}\Gamma$. Arrows indicate the time evolution and multi-branch behaviors are clearly shown. The panel (a) displays that quantum states with inversion of entanglement ordering are crossed (for example, points $\mathbf{A_1}$ and $\mathbf{A_2}$ where $C_1>C_2$ but $B_1<B_2$). The plots of the insets are related to the case of perfect cavity (single-mode reservoir). No one-to-one correspondence among couples of quantifiers exists in this case as well.}
\end{figure}

Further information can be found when examining the projections of
the whole curve on the $B$-$C$, $B$-$P$ and $C$-$P$ planes. We show
these projections in the case of a Lorentzian spectral distribution
having a width ten times larger than that in Fig.~\ref{fig:3D}. Such
a choice allows to distinguish more clearly the different curves.
All the panels of Fig.~\ref{fig:projections} show that there is no
one-to-one correspondence between any two of the quantities  $B$,
$P$ and $C$. It is interesting to notice that this behavior does not
depend on the losses of the cavity, but it remains true also when
the environment reduces to a single mode, as seen from the insets of
Fig.~\ref{fig:projections}. The absence of one-to-one correspondence
between any two of the quantities $C$, $P$ and $B$ is truly a
consequence of the reservoir-mediated interaction between the
qubits; in fact, if one examines the dynamics starting from the same
initial state, but with the two qubits embedded in independent
reservoirs, one-to-one correspondences between these quantities are
found. Considering the plot in the $B$-$C$ plane, displayed in panel
(a), it is possible to see that the system passes through states,
for example like those individuated by points $\mathbf{A_1}$ and
$\mathbf{A_2}$, such that $C_1>C_2$ but $B_1 < B_2$. This inversion
of entanglement ordering has been in general shown for different
quantifiers, as between entanglement of formation and either
negativity \cite{plenio1999JMO} or relative entropy of entanglement
\cite{miran2008PRA}. Indeed, there is a region characterized by
small values of concurrence ($0.30< C<0.35$) but where the Bell
inequality is violated up to values $\approx 2.1$. The $B$-$P$ plot
of panel (b) gives a justification of this behavior. In fact, as
already noticed from Fig.~\ref{fig:3D} in the $C$-$P$-$B$ space, to
these small values of concurrence, there correspond high values of
purity. In particular, when $P\approx0.95$ the maximum of Bell
function reaches $B\approx2.1$ (point $\mathbf{A_2}$). This
correspondence between small $C$ and high $P$ values is finally
confirmed by the $C$-$P$ plot of panel (c). Moreover, it is possible
to note that the system crosses the point $\mathbf{A_1}$ in the
$B$-$C$ plane two times (within the time interval we are
considering), in correspondence of which two different values of $P$
occur, as individuated by the points $\mathbf{A_1}$ and
$\mathbf{A_3}$ in the $C$-$P$ plane displayed in panel (c). This
means that at the same couple of values $C,B$, there correspond two
different values of $P$ ($P_1<P_3$). As a final remark, we note that
if one considers the part of plots where $B>2$ the multi-branch
behavior of Fig.~\ref{fig:3D} is retrieved.

For a lossy cavity, the analytic solution for the density matrix element is cumbersome as shown in Ref.~\cite{Mazzola2009ARX}. In the following we give these expressions in the simpler case when the common cavity has no losses.

\subsubsection{Perfect cavity}
For a lossless cavity (single mode reservoir) the density matrix
elements of the system can be expressed as function of the
population $\rho_{++}$ of the super-radiant state and of
$\rho_{14}$
\begin{equation}\label{perfect matrix}
  \rho_{11}=2|\rho_{14}|^2 , \quad  \rho_{22}=\rho_{33}=\rho_{23}=\rho_{++}/2.
\end{equation}
$\rho_{++}$ and $\rho_{14}$ are oscillating functions with different
periods (the first being the half of the second one) and their
expression is \cite{Mazzola2009ARX}
\begin{eqnarray}
\rho_{++} = \sin^2(\sqrt{6}\Omega t)/6 ,\quad
 \rho_{14} = [2+\cos(\sqrt{6}\Omega t)]/6,
\end{eqnarray}
where $\Omega$ is the coupling constant between the qubits and the
mode of the cavity. The interaction between the qubits, mediated by
the common reservoir, makes the coherence $\rho_{14}$ never vanish,
as instead happens in the case of independent reservoirs. From the
insets of Fig.~\ref{fig:projections}, one sees that there is not a
one-to-one correspondence between any two of the quantities $C$, $P$
and $B$. Due to the absence of losses in the cavity, the system goes
back and forth through the entire curve, meaning that at certain
times the qubits recover the pure maximally entangled state of
preparation. When dissipation is taken into account and a more
complex environment than a single mode is considered this picture
becomes more complex and the multi-way behavior in the $C$-$P$-$B$
parameter space of Fig.~\ref{fig:3D} and in the projection planes of
Fig.~\ref{fig:projections} arises.

The analysis above shows that a quite complex interplay occurs among
$C$, $P$ and $B$. It is known that in general, given two of these
quantities, this does not determine the third \cite{derkacz2005PRA}.
However, one may ask if it may happen that some explicit connections
among them exist, which can be expressed in a closed form for some
class of states.

\section{$C$-$P$-$B$ relation \label{sec-BPC}}
In this section we seek an equation among $C$, $P$ and $B$
that may be usefully adopted to quantify their connection in a
general context. To this purpose we shall generalize a relation
valid only for pure states. It is known that in this latter case, a
relation between $C$ and $B$ holds \cite{verstraete2002PRL},
\begin{equation}\label{CB relation}
 B=2\sqrt{1+C^2}.
\end{equation}
In the attempt to generalize this equation to mixed states, we
notice that the former equation can be written as $B=2\sqrt{P+C^2}$
with $P=1$. Therefore, it is rather natural to connect the three
quantities $C$, $P$ and $B$, for any state, as
\begin{equation}\label{BPCRformula}
B^2/4-P-C^2=R,
\end{equation}
where the ``remainder'' $R$ is a quantity expressed in terms of density
matrix elements that vanishes for pure states. In particular, four
different regions can be distinguished on the basis of $K_1,K_2$ and $u_2,u_3$ defined in Eqs.~(\ref{concurrenceandK}) and (\ref{uXstate}):
\begin{itemize}
  \item Region 1: $u_2\geq u_3$ and $K_{1}\geq K_{2}$
  {\setlength\arraycolsep{1.5 pt}\begin{eqnarray}\label{region1}
    B&=&B_1, \quad C=2 K_1,\quad R=R_1\nonumber\\
    R_1&=&2[|\rho_{23}|^2-|\rho_{14}|^2+\rho_{11}\rho_{44}-\rho_{22}\rho_{33}+4|\rho_{14}\rho_{23}|\nonumber\\
    &+&4|\rho_{14}|\sqrt{\rho_{22}\rho_{33}}-(\rho_{11}+\rho_{44})(\rho_{22}+\rho_{33})].
  \end{eqnarray}}
 \item Region 2: $u_2\geq u_3$ and $K_{2}\geq K_{1}$
  \begin{equation}\label{region2}
    B=B_1, \quad C=2 K_2,\quad R=R_2=R_1(1 \leftrightarrow 2, 3 \leftrightarrow 4).
    \end{equation}
  \item Region 3: $u_3\geq u_2$ and $K_{1}\geq K_{2}$
 {\setlength\arraycolsep{1.5 pt}\begin{eqnarray}\label{region3}
    B&=&B_2, \quad C=2 K_1,\quad R=R_3\nonumber \\
    R_3&=&2|\rho_{14}|^2+6|\rho_{23}|^2-4\rho_{22}\rho_{33}+8|\rho_{14}|\sqrt{\rho_{22}\rho_{33}} \nonumber\\
&-&\rho_{11}^2-\rho_{22}^2-\rho_{33}^2-\rho_{44}^2.
  \end{eqnarray}}
  \item Region 4: $u_3\geq u_2$ and $K_{2}\geq K_{1}$
  \begin{equation}\label{region4}
    B=B_2, \quad C=2 K_2,\quad R=R_4=R_3(1 \leftrightarrow 2, 3 \leftrightarrow 4),
  \end{equation}
\end{itemize}
where the symbol $i \leftrightarrow j$ means that index $i$ must be
changed into $j$ and viceversa. The introduction of a remainder in
Eq.~(\ref{BPCRformula}) allows us to express the Bell function as a
function of concurrence and purity and may explain why states
characterized by the same concurrence and purity can have different
values of the Bell function. Such a remainder might contain some unknown
properties qualifying the state of the system.

Even if in the general case a closed equation between $C$, $P$ and
$B$ does not exist, it may be useful to look for classes of states
for which the remainder can be expressed as a function of these same
quantities. In the following we show that this occurs in the case of
maximally entangled mixed states.

\subsection{\label{MEMSsection}Application to maximally entangled mixed states}
As an example to which to apply the considerations and the formulas
above we now consider the case of maximally entangled mixed states
(MEMS), defined as those states possessing the maximal amount of
entanglement (quantified by tangle $\tau$ or concurrence $C$) for a
given degree of mixedness (quantified by linear entropy $S$ or
purity $P$) \cite{munro2001PRA,wei2003PRA}. MEMS have been generated
in laboratory by parametric down conversion \cite{kwiat2004PRL};
their density matrix depends on the quantifiers chosen for
entanglement and mixedness. Typically, tangle $\tau=C^2$ is used to
quantify entanglement and linear entropy $S=\frac{4}{3}(1-P)$ to
quantify mixedness. Since the quantities $\tau$-$C$ and $S$-$P$ are
monotonically related each other, the use of $C$ and $P$ instead of
$\tau$ and $S$ does not affect the structure of MEMS density matrix.
For these quantifiers the explicit form of MEMS, in the standard
computational basis
$\mathcal{B}=\{\ket{11},\ket{10},\ket{01},\ket{00}\}$, is given (up
to local unitary transformations) by \cite{munro2001PRA}
\begin{equation}\label{MEMS}
\hat{\rho}_\mathrm{MEMS}= \left(
\begin{array}{cccc}
  g(\gamma) & 0 & 0 & \gamma/2  \\
  0 & 0& 0 & 0 \\
  0 & 0 & 1-2g(\gamma) & 0 \\
  \gamma/2 & 0 & 0 & g(\gamma) \\
\end{array}
\right),
\end{equation}
where the parameter $\gamma$ coincides with the concurrence $C$ (for any value of $\gamma $ the state is entangled) and
\begin{equation}
g(\gamma)=\left\{\begin{array}{lr}\gamma/2,& C\equiv\gamma\geq 2/3\\
1/3,& C\equiv\gamma<2/3\end{array}\right..
\end{equation}
According to the parametric regions identified by
Eqs.~(\ref{region1})-(\ref{region4}) and the $C$-$P$-$B$ relation of
Eq.~(\ref{BPCRformula}), we obtain the following expressions of $C$,
$P$, $B$ and $R$ for various ranges of $\gamma$:
\begin{itemize}
\item $0\leq\gamma\leq1/3$ corresponds to region 1 with
\begin{equation}
C=\gamma,\ P=\frac{1}{3}+\frac{\gamma^2}{2},\ \frac{B^2}{4}=\frac{1}{9}+\gamma^2,\ R=-\frac{2}{9}-\frac{\gamma^2}{2}.
\end{equation}

\item $1/3\leq\gamma\leq2/3$ corresponds to region 3 with
\begin{equation}
C=\gamma,\ P=\frac{1}{3}+\frac{\gamma^2}{2},\ \frac{B^2}{4}=2\gamma^2,\ R=-\frac{1}{3}+\frac{\gamma^2}{2}.
\end{equation}

\item $2/3\leq\gamma\leq1$ again corresponds to region 3 with
\begin{equation}\label{MEMSrest3}
C=\gamma,\ P=1-2\gamma+2\gamma^2,\ \frac{B^2}{4}=2\gamma^2,\ R=-(1-\gamma)^2.
\end{equation}
\end{itemize}
Regions 2 and 4 are excluded because for MEMS $K_1>K_2$ for any
value of $\gamma$. From the last three equations, it follows that
Bell inequality violation occurs only for $\gamma > 1/\sqrt{2}$. It
is worth to note that in this region of violation the maximum of
Bell function assumes the lower bound of violation,
$B_{\mathrm{low}}=2\sqrt{2}C$, for a given concurrence. This fact
can be considered as a further characterization of MEMS
\cite{verstraete2002PRL}. For any value of $\gamma$ the remainder $R$
results to be a function of only concurrence and vanishes when $P=1$
according to the considerations that follow the $C$-$P$-$B$ relation
of Eq.~(\ref{BPCRformula}). We recall that, by varying $\gamma$, the
MEMS individuate an upper bound curve in the $C$-$P$ plane under
which all the two-qubit quantum states are confined
\cite{munro2001PRA}.

In the following we come back to our dynamical case, showing that
choosing properly the initial state, the dynamics of the system
flows along this upper bound MEMS curve.

\section{Super-radiant state dynamics in $C$-$P$-$B$ space and MEMS generation}
\begin{figure}
\begin{center}
\includegraphics[width=6.5 cm, height= 4.5 cm]{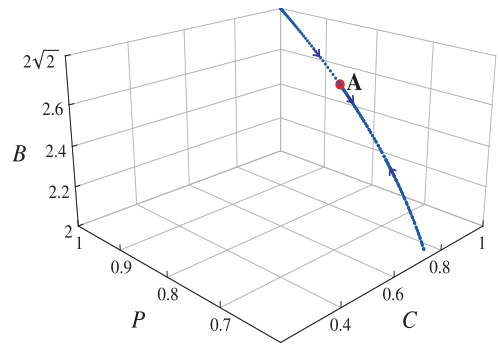}
\end{center}
\caption{\label{fig:3D1exc}\footnotesize (Color online) $C$-$P$-$B$ space curve drawn by the system starting from the initial one-excitation Bell state $\ket{+}$ for $\lambda=10^{-3}\Gamma$. The arrows indicate the time evolution and the point A the maximum point reached after the first ascent. A one-to-one correspondence among the three quantities is clearly shown.}
\end{figure}

Here, we investigate the dynamics of the two qubits in the same
model of Sec.~\ref{sec-model}, in the case they are initially
prepared in the one-excitation (super-radiant) Bell state
\begin{equation}\label{initial phi state}
    \ket{+}=(\ket{10}+\ket{01})/\sqrt{2}.
\end{equation}
The trajectory of the representative point of the system in the $C$-$P$-$B$ space is shown in Fig.~\ref{fig:3D1exc}.
This path is obtained by a dense sampling of triplets of $C$-$P$-$B$
values at different times (up to the time $200\Gamma t$). One sees
that the dynamics starts from the pure maximally entangled state
($C=1$, $P=1$), thus maximally violating the CHSH inequality
($B=2\sqrt{2}$). Due to the interaction with the environment, $C$,
$P$ and $B$ all decrease and at a certain time the CHSH inequality
is not violated anymore. After this time the curve goes below the
$B=2$ plane but after a while the memory effects of the
non-Markovian environment makes those three quantities
simultaneously revive. When the representative point raises above
the $B=2$ plane, giving revivals of $B$, it follows again the same
curve but runs only a part of it. This is related to the fact that
the system is open and environmental noise deteriorates the
coherence properties of the state of the system, with a
corresponding decrease of the maximum values of $C$ and $B$ with
time. Hence, the dynamics passes through cycles of revivals and
collapses until, after a certain time, the CHSH-Bell inequality is
not violated anymore.

This behavior can be clearly seen by examining the explicit
evolution of the two-qubit density matrix. All the density matrix
elements at a given time $t$ depend only on the population of the
super-radiant state $\rho_{++}$ at that time,
\begin{eqnarray}\label{phi density matrix}
   &&\rho_{11}=0,\quad \rho_{22}=\rho_{33}=\rho_{++}/2, \quad \rho_{44}=1-\rho_{++},\nonumber \\ &&
\rho_{23}=\rho_{++}/2,\quad\rho_{14}=0.
\end{eqnarray}
Varying the ratio between the spontaneous emission rate and the
spectral density width, $\Gamma/\lambda$, two different regimes in
the time behavior of $\rho_{++}$ can be distinguished. For
$\Gamma<\lambda/2$ (weak coupling) there is a Markovian exponential
decay controlled by $\Gamma$; for $\Gamma>\lambda/2$ (strong
coupling) non-Markovian effects become relevant. In this latter
regime the function $\rho_{++}$ assumes the form
\cite{maniscalco2008PRL}
\begin{equation}\label{p1}
\rho_{++}=\mathrm{e}^{-\lambda t}\left[ \cos \left(\frac{d
t}{2}\right)+\frac{\lambda}{d}\sin \left(\frac{d t}{2}\right)
\right]^2,
\end{equation}
where $d=\sqrt{2\Gamma \lambda-\lambda^2}$. In this strong coupling
regime $\rho_{++}$ presents damped oscillations while in the weak
coupling regime Markovian-like decay occurs (harmonic functions in
$\rho_{++}$ are replaced with the corresponding hyperbolic ones and
$d$ with $\imath d$). In the ideal cavity limit, $\lambda
\rightarrow 0$, $\rho_{++}$ becomes a purely oscillating function.

We point out that Eq.~(\ref{phi density matrix}) corresponds to the
density matrix form of MEMS of Eq.~(\ref{MEMS}) (for $\rho_{++} \geq
2/3$) where, after a local unitary transformation on one of the two
qubits (changing $\ket{0}$ in $\ket{1}$ and viceversa), $\rho_{++}$
plays the role of a time-dependent parameter $\gamma$, whose
behavior depends on the values of spectral density parameters. This
means that, starting from the super-radiant state, the two-qubit
system evolves along the MEMS curve. As a consequence, the physical
configuration of two qubits in a lossy common cavity is suitable for
a dynamical creation of MEMS (see also other proposals for MEMS
generation \cite{li2007PRA,campbell2009PRA}).

Because of Eq.~(\ref{p1}), clearly $C$, $P$ and $B$ do also depend
only on $\rho_{++}$. In particular for the range of values $0\leq
\rho_{++}\leq 1/3$ we are in the region 2 (see Sec.~\ref{sec-BPC})
and CHSH-Bell inequality is never violated; for
$1/3\leq\rho_{++}\leq 1$ we are in region 4 where $C$, $P$ and $B$
assume the form
\begin{equation}\label{Phi BPC}
    B=2\sqrt{2}\rho_{++},\quad
    P=1-2\rho_{++}(1-\rho_{++}),\quad
    C=\rho_{++},
\end{equation}
the CHSH-Bell inequality being violated for $\rho_{++}>1/\sqrt{2}$.
This form of $C$, $P$ and $B$ implies a closed relation among these
three quantities which can be analytically expressed as
\begin{equation}\label{BPCPhi}
B^2/4-P-C^2=-(1-C)^2,
\end{equation}
where the remainder $R$ of Eq.~(\ref{BPCRformula}) is given by
$R=-(1-C)^2$. Eq.~(\ref{BPCPhi}) corresponds to what obtained in
Eq.~(\ref{MEMSrest3}) for MEMS. It is worth to stress that,
differently from the general case where no closed relation among
$C$, $P$ and $B$ exists, here we deal with a dynamical case where a
closed relation is available. This analytical relation between $C$,
$P$ and $B$ explains why the system draws with time back and forth
on the same trajectory in the $C$-$P$-$B$ space. Moreover the
explicit expressions of Eq.~(\ref{phi density matrix}) allows to
understand why this trajectory remains unaltered when changing the
width of the Lorentzian distribution. Indeed, this is a consequence
of the fact that each $C$-$P$-$B$ point is determined by only one
specific value of $\rho_{++}$. In the case of a Lorentzian
distribution, $\rho_{++}$ exhibits damped oscillations between 0 and
1, so that repeated equal values of $\rho_{++}$ give the same
$C$-$P$-$B$ points and thus in turn the system dynamics draws back
and forth the same trajectory in the $C$-$P$-$B$ space. On the other
hand, the width of the Lorentzian affects the oscillatory behavior
of $\rho_{++}$, therefore influencing only the number of times and
how high the system can come back on the same curve in the
$C$-$P$-$B$ space. We emphasize once more that this is true only for
this particular initial state.

\section{Conclusions \label{sec-concl}}
In this paper the relation among entanglement, mixedness and
nonlocality in a two-qubit system has been investigated. The
nontrivial connection among the quantifiers of these properties,
namely concurrence $C$, purity $P$ and the maximum of Bell function
$B$ in the state space has been studied in a dynamical context. Two
qubits have been assumed to be embedded in a non-Markovian common
reservoir at zero temperature. Common reservoir-mediated interaction
and memory effects induce, with different intensities, revivals of
all the three quantities. The $C$-$P$-$B$ ``parameter'' space has
been introduced and exploited for the description of the relations
among $C$, $P$ and $B$ for the two-qubit reduced dynamics.

For an initial two-excitation Bell state, it has been shown that the
system draws a multi-branch curve in the $C$-$P$-$B$ space.
Projection of this curve on two-dimensional spaces clearly shows the
absence of one-to-one correspondence between couples of the
quantifiers $C$, $P$ and $B$ \cite{derkacz2005PRA}. This dynamical feature is maintained even in
the limit of perfect cavity suffering no losses. A comparison with
the case of independent reservoirs, where this correspondence
between couples of the quantifiers occurs, has been made evidencing
the role of the common reservoir-mediated interaction between qubits
as responsible of the lack of such correspondence.

The search of classes of states where a closed relation among $C$,
$P$ and $B$ holds, has led us to look for general connections among
these quantifiers. On the basis of known relations between
concurrence and maximum of Bell function in the pure state case, an
extended relation between all the three quantifiers for a wide class
of mixed states has been given. A remainder, vanishing in the limit of
pure state, has been introduced and its explicit form given for four
different regions identified by the quantum state under
investigation. This term could play a role to explain the
complex and not well understood relation among all these quantities.
Moreover we have shown that, for the class of maximally entangled
mixed states (MEMS), a closed relation among $C$, $P$ and $B$
exists.

In the final part of the paper we have reconsidered our dynamical
model, showing that if the two qubits are initially prepared in the
one-excitation Bell state (super-radiant state), differently from
the two-excitation case, a one-to-one correspondence between any
couple of $C$, $P$ and $B$ occurs. This results in a single-valued
relation represented by a one-branch curve in the $C$-$P$-$B$ space
which is drawn back and forth by the system. In this case we have a
physical configuration in which a closed analytical relation among
$C$, $P$ and $B$ can be written. We have moreover shown that the
system evolves maintaining the MEMS density matrix structure.
Therefore this physical configuration may be seen as a suitable
setup for MEMS generation.

\begin{acknowledgments}
L.M. thanks S. Maniscalco and J. Piilo for useful discussions, and G.
Compagno and his group for the kind hospitality at the Universit\`{a} di Palermo.
L.M. thanks also M. Ehrnrooth Foundation for financial support.
\end{acknowledgments}

\end{document}